\pdfoutput=1

\documentclass[twocolumn,aps,prl,superscriptaddress]{revtex4-1}

\usepackage{amssymb,amsfonts,amsmath,bm}
\usepackage{color,graphicx}
\usepackage[normalem]{ulem}

\newcommand{\R}{\mathbb R}

\newcommand{\m}{\mathbf{m}}
\newcommand{\n}{\hat{\mathbf{n}}}

\newcommand{\beml}{\begin{subequations}}
\newcommand{\eml}{\end{subequations}}

\begin{document}

\title{Walker solution for Dzyaloshinskii domain wall in ultrathin ferromagnetic films}
 
\author{Valeriy V. Slastikov} \affiliation{School of Mathematics, University of Bristol, Bristol BS8 1TW, United Kingdom}
  
\author{Cyrill B. Muratov}
\affiliation{Department of Mathematical Sciences, New Jersey Institute of Technology, Newark, New Jersey 07102, USA} 

\author{Jonathan~M. Robbins}
\affiliation{School of Mathematics, University of Bristol, Bristol BS8 1TW, United Kingdom}
  
\author{Oleg A. Tretiakov}
\email{o.tretiakov@unsw.edu.au}
\affiliation{School of Physics, The University of New South Wales, Sydney 2052, Australia}
\affiliation{Institute for Materials Research and Center for Science and Innovation in Spintronics, Tohoku University, Sendai 980-8577, Japan}
\affiliation{National University of Science and Technology MISiS, Moscow 119049, Russia}


\begin{abstract}
  We analyze the electric current and magnetic field driven domain
  wall motion in perpendicularly magnetized ultrathin ferromagnetic
  films in the presence of interfacial Dzyaloshinskii-Moriya
  interaction and both out-of-plane and in-plane uniaxial
  anisotropies. We obtain exact analytical Walker-type solutions in
  the form of one-dimensional domain walls moving with constant
  velocity due to both spin-transfer torques and out-of-plane magnetic field. These solutions are embedded into a larger family of
  propagating solutions found numerically. Within the considered
  model, we find the dependencies of the domain wall velocity on the
  material parameters and demonstrate that adding in-plane anisotropy
  may produce domain walls moving with velocities in excess of 500 m/s
  in realistic materials under moderate fields and currents.
\end{abstract}

\maketitle

\textit{Introduction.} In their seminal paper, Schryer and Walker discovered an exact
analytical solution of the Landau-Lifshitz-Gilbert (LLG) equation
describing a moving one-dimensional (1D) domain wall (DW)
\cite{Walker74}. In this so-called Walker solution, the magnetization
rotates in a fixed plane determined by the material parameters and
magnetic field, connecting the two opposite in-plane equilibrium
orientations of magnetization. The Walker solution has since been used
in numerous problems of DW motion to successfully explain the physics
of magnetization reversal
\cite{Atkinson03,Yamaguchi04,Allwood02,Parkin:racetrack08,Tatara04,Duine07,Tretiakov08,BeachPRL09,Krivorotov10,Tretiakov_DMI,Fert2013,Shibata2011_review,Hoffmann2015_review}.

Recently, out-of-plane magnetized ultrathin films with
Dzyaloshinskii-Moriya interaction (DMI)
\cite{Dzyaloshinskii58,Moriya60} have attracted significant interest
\cite{Thiaville2012,Boulle13,Beach2013,Parkin2013,Brataas2013,Ohno2014,Emori2013,Franken2014,
  Martinez2014,Vandermeulen2016,Jiawei2016} due to their potential
advantages for high-performance spinorbitronic devices
\cite{Miron2011,Emori2013,Ono2016}. These materials are known to
exhibit chiral DWs \cite{Ono2016,Panagopoulos2016review,Garg2018}, but
so far no explicit dynamic Walker-type solution has been
demonstrated to exist, which significantly hinders understanding of
the DW motion in these systems.

In this Rapid Communication, we report a new exact analytical solution for steady
DW motion in out-of-plane magnetized films analogous to the Walker
solution for films with in-plane equilibrium magnetization. For this
solution to exist, a small in-plane anisotropy is required in addition
to the dominant out-of-plane anisotropy, while the film is still
magnetized out-of-plane. We consider both current and field driven DW
dynamics in the presence of interfacial DMI and show that this new
solution can describe the DW motion observed in recent experiments
\cite{Emori2013,Franken2014,Jiawei2016}.

At nonzero DMI strength, our solution fixes the angle of magnetization
in the DW such that it acquires a strictly N\'eel profile. The
solution also fixes the angle between the direction of the current and
the DW normal. This angle depends on the relative strength of magnetic
field and electric current, but, notably, is independent of the DMI
strength. Moreover, in the absence of DMI we find an entire family of
exact solutions for every angle between the DW normal and the in-plane
easy axis. Although the dynamics in biaxial ferromagnets has been the
subject of many works (see, e.g., \cite{Kosevich1990,Thiaville05,
  goussev13, Su2017, nasseri18}), the interplay between DMI and biaxial
anisotropy leads to additional interesting phenomena.

\begin{figure}[t] \centering
  \includegraphics[width=3.4in]{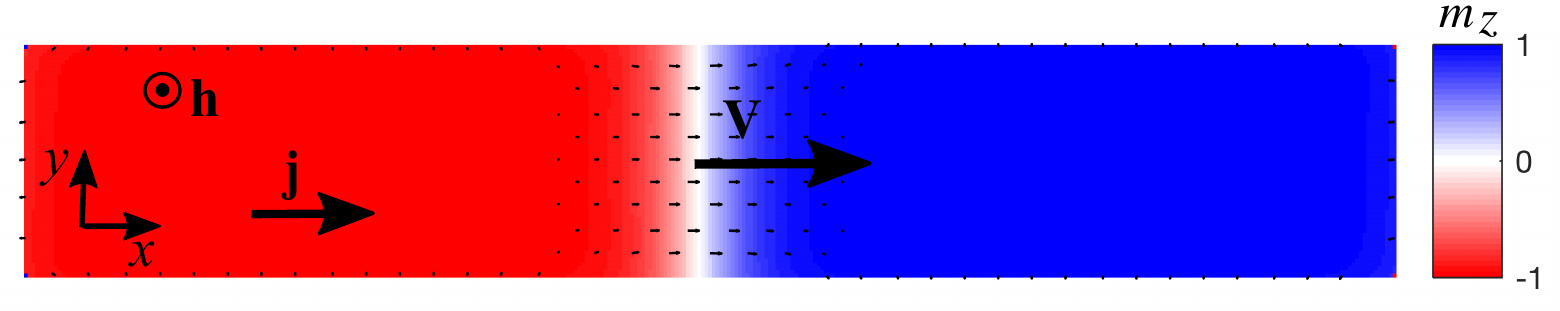} \caption{A
    snapshot of tiltless DW driven by current $\mathbf{j}$ and
    out-of-plane magnetic field $\mathbf{h}$ in a ferromagnetic
    nanostrip with DMI and two anisotropies (larger out-of-plane and
    smaller in-plane along the strip axis) from a simulation of
    Eq.\,\eqref{LLG}. }
\label{fig:DW_sketch}
\end{figure}

We also demonstrate that one can achieve the highest propagation
velocities for tiltless DWs, i.\,e., DWs which move along the current
with the DW front strictly perpendicular to the current
(Fig.\,\ref{fig:DW_sketch}), by appropriately tuning the magnetic
field. As a result, we provide an exact experimentally
relevant~\cite{Boulle13,Emori2013,Franken2014,Jiawei2016} way to
achieve the \textit{maximal} DW velocity in a nanowire for a given
current. We note that in thin nanowires, the direction of current
along the wire coincides with the direction of the in-plane easy-axis
shape anisotropy due to stray fields \cite{kohn05arma}.

\textit{Model.} We consider an ultrathin ferromagnetic film of
thickness $d$ with interfacial DMI and two anisotropies: larger
out-of-plane and smaller in-plane, and study the dynamic behavior of
magnetic DWs due to an out-of-plane magnetic field and/or in-plane
electric current. Our analysis is based on the LLG equation with
spin-transfer torques \cite{Tatara04,Thiaville05} describing the
evolution of the reduced magnetization $\mathbf m(\mathbf r, t)$
\cite{suppl}: \begin{equation}
  \label{LLG}
  \frac{\partial \mathbf{m}}{\partial t}
  =   \mathbf{h}_{\rm{eff}} \times \mathbf{m} + \alpha \mathbf{m}\times
  \frac{\partial \mathbf{m}}{\partial t} - (\mathbf{j} \cdot \nabla)
  \m + \beta \m \times  (\mathbf{j} \cdot \nabla)
  \m , 
\end{equation}
where $\mathbf r \in \mathbb R^2$ is the spatial coordinate in units
of the exchange length $\ell_{ex} = \sqrt{2 A / (\mu_0 M_s^2)}$ and
$t$ is time in the units of $(\gamma \mu_0 M_s)^{-1}$, $A$ is the
exchange stiffness, $M_s$ is the saturation magnetization, $\gamma$ is
the gyromagnetic ratio, $\alpha$ is the Gilbert damping constant,
$\beta$ is the nonadiabatic spin-transfer torque constant,
$\mathbf j = \hbar P \mathbf J / \sqrt{8 A e^2 \mu_0 M_s^2}$, 
$\mathbf{J}$ is the in-plane current density, $P$ is the spin
polarization of current, and
$\mathbf h_\textrm{eff} = - \delta E /\delta \m$ with energy $E$ in
units of $2Ad$ given by
 \begin{align}
   \label{Eh}
   \!\!E(\m)  = & \frac12 \int_{\R^2} \Big[ |\nabla \m|^2 +
          (k_z - 1) (1-m_z^2) - k_x m_x^2 
          \notag \\
        & - 2 {h_z} m_z  + \kappa \left( m_z \nabla \cdot \m' -
          \m' \cdot \nabla m_z \right) \Big] \, d^2 r.
 \end{align}
Here $\m = (\m', m_z)$, $\m' = (m_x, m_y)$, and we introduced
 the dimensionless parameters corresponding to the dimensional
 out-of-plane and in-plane anisotropy constants $K_z$ and $K_x$,
 interfacial DMI constant $D$, and out-of-plane field $H_z$, respectively:
 \begin{align}
  \label{Qkappa}
   k_{x,z} = {2 K_{x,z} \over \mu_0 M_s^2},\quad \kappa = D \sqrt{2
   \over \mu_0 
  M_s^2 A}, \quad h_z = {H_z \over M_s}. 
\end{align}
We assume $k_z > 1$ and $0 < k_x < k_z - 1$ to ensure that
$\m = \pm \hat{\mathbf z}$ are the only stable equilibria for
$h_z = 0$. The energy in Eq.\,\eqref{Eh} is appropriate for ultrathin
films, i.\,e., for $d / \ell_{ex} \ll 1$ \cite{Muratov2017}. Note that
Eq.~\eqref{LLG} does not include spin-orbit torques, which may be
important in bilayer/multilayer ferromagnetic structures with
heavy-metal layers, where electric currents run in the presence of
strong spin-orbit interaction
\cite{Garello2013,Ado2017,Manchon2018,Yang2018:SOTreview}. However,
spin-orbit torques affect not just the DW itself, but the entire
magnetization configuration in the film, thus precluding the existence
of Walker-type solutions.

\textit{DW profile.} We study the dynamics of DWs moving due to either an applied magnetic field or a spin-transfer torque from an electric current.  By a moving DW with normal velocity $V$ in the direction of the unit vector $\n = (n_x, n_y)$ we mean a 1D solution of~\eqref{LLG} of the form $\m = \m (\mathbf r \cdot \n - Vt)$. Substituting this traveling wave ansatz into Eq.\,\eqref{LLG} and writing
$\m = (\sin \theta \cos \phi , \sin \theta \sin \phi, \cos \theta)$ yields the following system of differential equations for $\theta$ and $\phi$ as functions of $\xi = \mathbf r \cdot \n - Vt$~\cite{suppl}:
\begin{align}
\label{LLG_final1}
  &\frac{1}{\sin\theta} \frac{d}{d \xi}\!\left(\!
    \sin^2\theta  \frac{d \phi}{d \xi}\!\right)\! 
    + \! (\alpha  V \! - \! \beta \mathbf j \cdot \n)  \sin\theta \frac{d \phi}{d \xi}
    \nonumber\\
  &\! - \! (\mathbf{j}\cdot \n \! - \! V \! + \! \kappa \n \cdot \hat{\mathbf{p}}
    \sin\theta ) \frac{d \theta}{d \xi}   
    - \frac{k_x}{2} \sin\theta \sin2\phi  =0, \\
\label{LLG_final2}
  & \frac{d^2 \theta}{d \xi^2}  \! + \! (\alpha V \! - \! \beta
    \mathbf{j}\cdot \n )\frac{d \theta}{d \xi}   
    \!+ \! ( \mathbf j \cdot \n \! - \! V \! + \! \kappa \n\cdot
    \hat{\mathbf{p}} \sin\theta )\sin \theta \frac{d \phi}{d \xi} 
    \nonumber \\ 
  &  \! - \! \left(\! k_z \! - \! 1+ \! \left| \frac{d \phi}{d
    \xi}\right|^2 \hspace{-1mm} \! - \! k_x
    \cos^2\phi \!\right)\! \sin \theta \cos \theta 
    \! - \! h_z \sin\theta  \! = 0, \! 
\end{align}
where for convenience we defined $\hat{\mathbf{p}}= (-\sin \phi, \cos \phi)$. Equations
(\ref{LLG_final1}) and~(\ref{LLG_final2}) need to be supplemented by the
conditions at infinity. With the convention that the positive velocity
($V > 0$) corresponds to a domain with $\m = -\hat{\mathbf z}$
invading the domain with $\m = \hat{\mathbf z}$, we require
$\theta(-\infty) = \pi$ and $\theta(+\infty) = 0$. The DW velocity $V$
is determined by solvability of
Eqs.~\eqref{LLG_final1} and~\eqref{LLG_final2}.

\textit{Walker solution.} In the absence of DMI ($\kappa =0$),
Eqs.~\eqref{LLG_final1} and~\eqref{LLG_final2} admit an exact solution
for every $\n$ with the help of the Walker ansatz~\cite{Walker74},
thereby generalizing the results of Ref.~\cite{goussev13} to
two-dimensional (2D) film. Namely, setting
$\phi =\phi_0 =\mathrm{const}$ and matching the second derivative of
$\theta(\xi)$ to the term proportional to $\sin 2 \theta$ yields
\begin{align} \label{eq1}
  h_z \sin \theta - (\alpha V - \beta \mathbf j \cdot \n) \theta'
  &=0, \\  
  \label{eq2}
 \theta'' - (k_z-1 - k_x \cos^2 \phi_0)
  \sin\theta \cos\theta &=0, 
  \\ 
\label{eq3}
  (V - \mathbf j \cdot \n) \theta' - \tfrac12 k_x
  \sin\theta\sin2\phi_0 &=0, 
\end{align}
where $\theta' = d \theta / d \xi$ and $\theta'' = d^2 \theta / d \xi^2$. This system of equations produces a Walker-type solution for a steadily moving DW:
\begin{equation}
  \label{eq:walkersol}
  \theta(\xi) = 2 \arctan e^{-\xi \sqrt{k_z - 1 - k_x \cos^2 \phi_0}},
\end{equation}
propagating with velocity
\begin{equation}
  \label{eq:V}
  V = - {h_z \over \alpha \sqrt{k_z - 1 - k_x \cos^2 \phi_0}} + {\beta
    \mathbf j \cdot \n \over \alpha},
\end{equation}
where $\phi_0$ solves
\begin{equation}
  \label{eq:n}
  \mathbf j \cdot \n (\alpha \! - \! \beta) \sqrt{k_z \! - \! 1 \! -
    \! k_x \cos^2
    \phi_0}
  \! + \! h_z \! = \! \tfrac12 \alpha k_x \sin 2 \phi_0.
\end{equation}
The obtained front velocity depends on the propagation direction $\n$,
unless $\mathbf j \cdot \n = 0$. In particular, at $h_z = 0$ the velocity is maximal
in the direction of $\mathbf j$. The solution exists only when $|h_z|$
and $j = |\mathbf j|$ do not exceed critical values corresponding to
Walker breakdown \cite{Walker74,Clarke08,goussev13}.

In the presence of DMI ($\kappa \not= 0$) the Walker solution obtained above is generally
destroyed. Nevertheless, Eqs.~\eqref{eq1}--\eqref{eq3} are preserved
in the special case when $\phi_0$ is chosen so that $\n\cdot \hat{\mathbf{p}} =0$. This
condition is equivalent to
\begin{align}
  \label{eq:nsign}
  \n = \pm (\cos \phi_0, \sin \phi_0),   
\end{align}
corresponding to a N\'eel-type DW profile, in which the magnetization
rotates entirely in the $\n$-$\hat{\mathbf z}$ plane. We stress that
Eq.~\eqref{eq:nsign} is dictated by solvability of
Eqs.~\eqref{eq1}--\eqref{eq3} and is not an assumption. In terms of
the space-time variables, the solution is given by
\begin{align}
  \label{eq:mwalker}
  \m(\mathbf r, t) = (\pm \n \sin \theta(\mathbf r \cdot \n - Vt), \cos
  \theta(\mathbf r \cdot \n - Vt)),    
\end{align}
where $\theta$ is given by Eq.~\eqref{eq:walkersol}, and ``$\pm$''
corresponds to the  choice of the sign in Eq.~\eqref{eq:nsign}. This
is an exact  Walker-type solution valid in the presence of
interfacial DMI and describing a 1D moving DW. Its propagation
direction is given by Eq.~\eqref{eq:nsign} in which
$\phi_0$ solves 
\begin{align}
  \label{eq:nn}
  h_z - \tfrac12 \alpha k_x \sin 2 \phi_0 \pm  (\alpha - \beta)  (j_x
  \cos \phi_0 + j_y \sin \phi_0) \notag \\ 
  \times \sqrt{k_z - 1 - k_x \cos^2 \phi_0}  
  = 0,
\end{align}
for $\mathbf j = (j_x, j_y)$, according to Eq.~\eqref{eq:n}. In
general, Eq.~\eqref{eq:nn} reduces to a fourth-order equation in
$\cos^2 \phi_0$, whose roots can in principle be found for all
parameters. Below we consider two important cases of purely current or
field driven DW motion, which are simpler mathematically and contain
all the essential physics.

Before concentrating on moving DWs, we consider the case of no applied
field and current, corresponding to static DWs (for further
  details, see, e.g., Ref.~\cite{Rohart2013}).  With $h_z =j =0$,
Eq.~\eqref{eq:nn} yields four distinct solutions:
$\phi_0 = -{\pi \over 2}, 0, {\pi \over 2}, \pi$. Then, inserting the
profile from Eq.~\eqref{eq:mwalker} with $V = 0$ into Eq.~\eqref{Eh},
one obtains the static DW energy per unit length \begin{equation}
  \label{eq:E0}
  E_0  = 2 \sqrt{k_z - 1 - k_x \cos^2 \phi_0} \mp \tfrac12 \kappa \pi.
\end{equation}
The DW energy $E_0$ is positive and is minimized by $\phi_0 = 0, \pi$
for $|\kappa| < (4/\pi) \sqrt{k_z - 1 -k_x}$. Furthermore, the DMI
contribution is minimized by the ``$+$'' sign in Eq.~\eqref{eq:nsign}
when $\kappa > 0$, and by the ``$-$'' sign when $\kappa < 0$. These
minimizing choices of $\phi_0$ and the sign in Eq.~\eqref{eq:nsign}
yield global minimizers (up to translations) of the 1D DW energy under
the conditions $\theta(-\infty) = \pi$ and $\theta(+\infty) = 0$ for
Eqs.\,(\ref{LLG_final1}) and~(\ref{LLG_final2}), since in this case both
the DMI and the in-plane anisotropy energy contributions are
separately minimized~\cite{Muratov2016}. Thus, the choices of $\n$
dictated by Eq.~\eqref{eq:nsign} with the above choices of $\phi_0$
and the sign correspond to the DW orientations with the lowest $E_0$.

We now consider two characteristic cases of moving DWs. For
definiteness, we assume $\kappa > 0$ and fix the positive sign in
Eq.~\eqref{eq:nsign}, corresponding to the minimum of the static DW
energy. It then allows us to think of $\phi_0$ as the angle defining
the normal vector in the direction of DW propagation whenever $V > 0$.
In the simplest case of no current, we find that for
$|h_z| \leq h_z^c$ the propagation angle of a DW solving
Eq.~\eqref{eq:nn} satisfies \begin{equation}
  \label{hzkxphi0}
  \sin 2 \phi_0 = 2h_z/(\alpha k_x), \qquad h_z^c =
  \alpha k_x / 2. 
\end{equation}
Once again, this equation produces four distinct values of
$\phi_0 \in (-\pi, \pi]$ for $|h_z|$ below the Walker breakdown field
$h_z^c$. Due to the symmetry $\phi_0 \to \phi_0 +\pi$, $\n \to -\n$
for $j=0$, this still results in two distinct solutions (differing by
180$^\circ$ rotations) with propagation velocities determined by
Eq.~\eqref{eq:V}. For both values, the sign of $V$ coincides with that
of $-h_z$, while the magnitude of $V$ is maximized by
$\phi_0 = \tfrac12 \arcsin [2 h_z /(\alpha k_x)]$.  This choice
corresponds to the branch of solutions that connects to the global DW
energy minimizers as $h_z \to 0$ and should thus correspond to the
physically observed solution. The DW velocity is
\begin{equation}
  \label{eq:Vhz}
  V = - \frac{h_z}{\alpha \sqrt{k_z-1 - \frac{k_x}{2} \left( 1 
        + \sqrt{1 - {4 h_z^2 \over \alpha^2 k_x^2 }} \, \right)}}.
\end{equation}
In particular, the velocity $V$ and angle $\phi_0$ at small fields
grow linear in $h_z$, while for $|h_z|$ comparable to $h_z^c$ they
acquire a nonlinear character. The magnitude of $|\phi_0|$ is a
monotonically increasing function of $|h_z|$, whose maximum
$|\phi_0| = \pi/4$ is achieved at the Walker breakdown field
$|h_z| = h_z^c$. Also, the DMI part of the DW energy is, in fact,
globally minimized by our sign choice in Eq.~\eqref{eq:nsign}.

Next, we study the case of purely current driven DW motion with
$\mathbf j = (j_x, 0)$ along the in-plane easy axis. By
Eq.~\eqref{eq:nn} one DW solution corresponds to a profile with
$V = 0$ and $\phi_0 = \pm \pi/2$. For $|j_x| < j_x^c$, where the
critical "Walker breakdown" current is \begin{equation}
  \label{eq:jzc}
  j_x^c = \alpha k_x /(|\alpha - \beta| \sqrt{k_z -1}),
\end{equation}
Eq.~\eqref{eq:nn} has two additional solutions:
\begin{equation}
  \label{eq:phi0j}
  \phi_0 = \arcsin \left( {(\alpha - \beta) j_x \over k_x} \sqrt{\frac{k_x( k_z
  -1 - k_x)}{ \alpha^2 k_x - j_x^2\left( \alpha - 
    \beta \right)^2}} \, \right), 
\end{equation}
and another one obtained by changing $\phi_0 \to \pi - \phi_0$ (and $V
\to -V$ in the  equation for the velocity). Focusing on the first
solution and substituting  the angle from Eq.~\eqref{eq:phi0j} into
Eq.~(\ref{eq:V}), we  obtain
\begin{equation} 
  V = \frac{\beta  j_x}{\alpha} \sqrt{\alpha^2 k_x^2 - j_x^2 (\alpha -
    \beta)^2 (k_z - 1) \over \alpha^2 k_x^2 - j_x^2 (\alpha -
    \beta)^2 k_x}.
\end{equation}
In the purely current driven case the DW velocity in the horizontal
direction $V_x = V / \cos \phi_0$ takes a universal form
$V_x = \beta j_x / \alpha$ [see Eq.~\eqref{eq:V}] also found for
current-induced DW and skyrmion motion in other systems
\cite{Thiaville05,goussev13,Goussev2016,Barker2016}. In particular,
the DW is driven only by the non-adiabatic torque. As $j_x$ is
increased, the angle $\phi_0$ monotonically increases, first linearly
in $j_x$ and then acquiring a nonlinear character closer to its
maximum $|\phi_0| = {\pi \over 2}$ at $|j_x| = j_x^c$. For larger
currents one would expect $|\phi_0|$ to remain equal to $\pi/2$,
consistent with the above static DW solution.

\textit{Other traveling-wave solutions.} As we just demonstrated, the
Walker-type solutions obtained for $\kappa \not= 0$ exist only for
certain specific directions of propagation determined by the solutions
of Eqs.~\eqref{eq:nn} and \eqref{eq:nsign}. In contrast, for
$\kappa=0$ there exists a traveling wave solution for every direction
$\n$, provided that $h_z$ and $j$ are not too large. To investigate
this further, we carried out numerical simulations of the 1D version
of Eq.~\eqref{LLG} with initial condition
$\m(\mathbf r, 0) = (\n \, \textrm{sech}(\frac12 \mathbf r \cdot \n),
\tanh (\frac12 \mathbf r \cdot \n))$, where $\n$ is given by
Eq.~\eqref{eq:nsign} with the ``+'' sign, and determined the long-time
asymptotic DW profile. For all parameter choices used in our
simulations the solution always converged to a DW moving with a
constant velocity $V = V(\phi_0)$. In particular, for every
propagation direction we found a propagating DW solution, which
coincided with the Walker-type solution obtained above for the
particular propagation direction satisfying Eq.~\eqref{eq:nn}. We
illustrate our findings with simulation results for the material
parameters as in \cite{Boulle13}: $A = 10^{-11}$ J/m,
$M_s = 1.09 \times 10^6$ A/m, $K_z = 1.25 \times 10^6$ J/m$^3$,
$D = 1$ mJ/m$^2$, and $\alpha = 0.5$.

\begin{figure}
  \centering
  \includegraphics[width=3.2in]{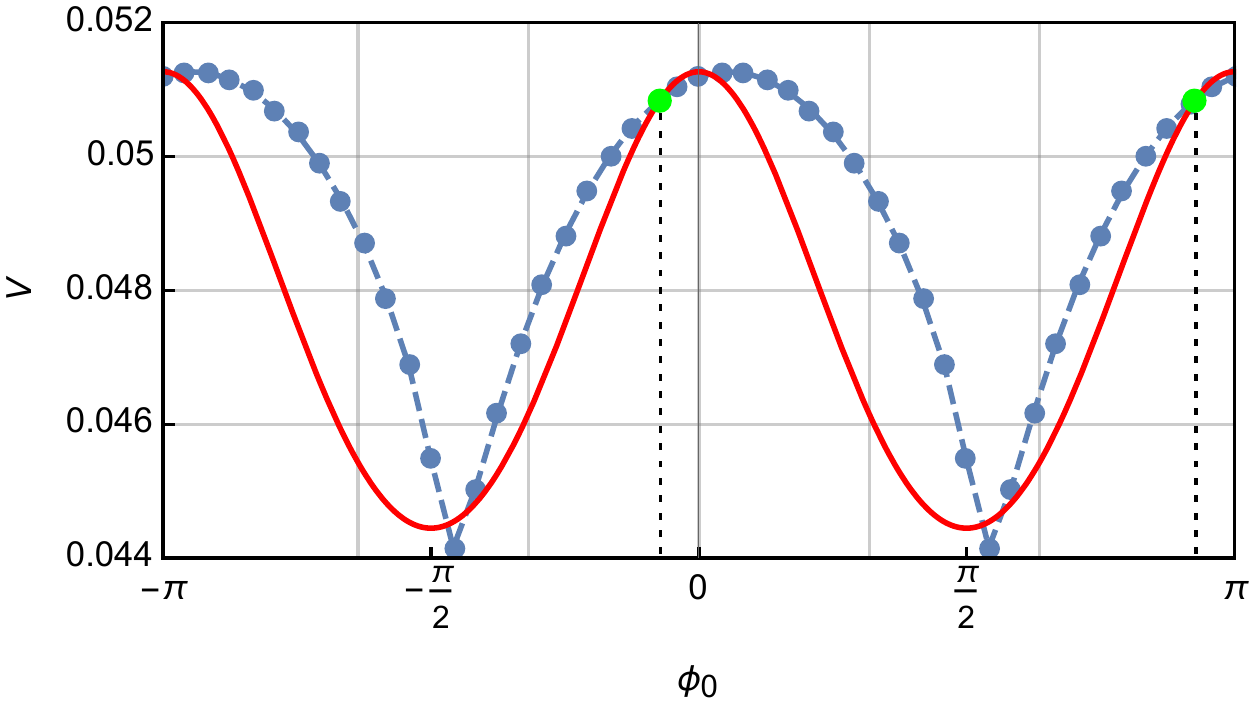}
  \caption{The dimensionless DW velocity $V$ at zero current as a
    function of the propagation angle $\phi_0$ obtained from the
    solution of Eq.~\eqref{LLG} for $k_z = 1.674$, $k_x = 0.167$,
    $\kappa = 0.366$, $\alpha = 0.5$, and $h_z = -0.0183$
    corresponding to the parameters in the text. The simulated data
    are indicated by the blue dots connected with a dashed blue
    line. The red solid line shows the dependence given by
    Eq.~\eqref{eq:V} with $\mathbf j = 0$. The green dots indicate the
    velocity for the Walker solution from Eq.~\eqref{eq:Vhz},
    corresponding to the special values of $\phi_0$ obtained from
    Eq.~\eqref{hzkxphi0} and indicated by dotted lines.}
  \label{fig:Vvsphi0}
\end{figure}

With no current, we carried out simulations for in-plane anisotropy
constant $K_x = 0.125 \times 10^6$ J/m$^3$ and applied field
$\mu_0 H_z = -25$ mT, corresponding to $|h_z|$ comparable to the
Walker breakdown field $h_z^c$ and a relatively small
$k_x$~\cite{suppl}. We then obtained the DW profile and velocity as
functions of propagation direction. The profile was found to be close
to that of the Walker solution, coinciding with it exactly when
$\phi_0$ solves Eq.~\eqref{eq:nn}. A plot of $V(\phi_0)$ is presented
in Fig.~\ref{fig:Vvsphi0}, indicating the points corresponding to the
Walker solution with green dots.

For small values of $k_x$ the DW moves with velocity nearly
independent of direction and its magnitude is close to the velocity of
the Walker solution. In this case the DW velocity's dependence on
propagation angle, $V(\phi_0)$, is well approximated by
Eq.~\eqref{eq:V}. On the other hand, as the value of $k_x$ is
increased, the velocity begins to exhibit a substantial dependence on
propagation angle and deviates from the prediction of
Eq.~\eqref{eq:V}, except for the Walker solution, even if the latter
still gives a fairly good approximation to its magnitude. When $k_x$
approaches its maximum value of $k_z - 1$ the velocity exhibits a
strong directional dependence that is not captured by
Eq.~\eqref{eq:V}, except, once again, for the Walker solution. Note
that the original dimensional propagation velocity
$V \sqrt{2 A \mu_0 \gamma_0^2}$ reaches $\sim\!500$ m/s. Thus, the
effect of a large in-plane uniaxial anisotropy is to accelerate the DW
by promoting the magnetization rotation in the easy in-plane
direction.

Similar results were obtained for current driven DW motion with no
applied field. For example, for $K_x = 0.4 \times 10^6$ J/m$^3$,
$\beta = 0.25$, $P = 1$, and $J_x = 5 \times 10^{12}$ A/m$^2$, we
found that the DW velocity is given by Eq.~\eqref{eq:V} with
$h_z = 0$. This is consistent with the expected physical picture that
the DW is advected with the velocity $V_x = \beta j_x / \alpha$ along
the current direction.

\textit{Motion along the in-plane easy axis.} The analysis of the
Walker solution performed above indicates that one can also select the
Walker solution moving in a {\em prescribed} direction given by angle
$\phi_0$ via an appropriate choice of the relationship between $h_z$
and $\mathbf j$. Furthermore, according to Eq.~\eqref{eq:V}, for fixed
$h_z < 0$ and $j$ the maximum velocity of the Walker solution is
achieved for $j_y =\phi_0 =0$. Substituting this into
Eqs.~\eqref{eq:nn} and \eqref{eq:V} then yields \begin{align}
  \label{eq:Vmax}
  V = j_x \quad \text{for} \quad h_z = j_x (\beta -
  \alpha) \sqrt{k_z - 1 - k_x}.
\end{align}
This maximal velocity turns out to be independent of most of the
material parameters, and the required field $h_z$ vanishes in the
special case $\alpha =\beta$. Furthermore, these solutions correspond
to moving DWs with no tilt, contrary to those seen in
Ref.~\cite{Boulle13} without in-plane anisotropy.

\textit{Traveling waves for zero damping.} It is interesting that the
obtained Walker solution also allows one to construct steadily moving DW
solutions at zero damping, $\alpha = 0$, for any angle $\phi_0$ by
taking the limit $\alpha \to 0$, while choosing $h_z$ to satisfy
Eq.~\eqref{hzkxphi0} with $\mathbf j = 0$. Substituting this into
Eq.~\eqref{eq:V} yields yet another exact solution valid for
$j =h_z = \alpha = 0$, in the form of a DW moving with velocity
\begin{align}
  \label{eq:V0}
  V = -{k_x \sin \phi_0 \cos \phi_0 \over \sqrt{k_z - 1 - k_x \cos^2
  \phi_0}},  
\end{align}
in the direction of $\n$ in Eq.~\eqref{eq:nsign} and with profile given by Eq.~\eqref{eq:mwalker}. This solution represents a 1D solitary wave propagating in the direction characterized by $\phi_0$ in the Hamiltonian setting, in the presence of interfacial DMI.

\textit{2D simulations.} To illustrate the role of the obtained DW solutions in magnetization reversal, we carried out full numerical simulations of Eq.~\eqref{LLG} in a nanostrip. The onset of a tiltless DW propagation due to both current and out-of-plane field is given in the Supplemental
Material movie \footnote{See Supplemental Material at [URL to be inserted by publisher] for the movie demonstrating the DW propagation due to both applied field and electric current using micromagnetic simulations.}. A snapshot of the steadily moving DW from this simulation is shown in Fig.~\ref{fig:DW_sketch}. We used the same parameters as in 1D simulations above~\cite{suppl}. The initial state was a single N\'eel DW across the strip at $j=h_z =0$. In the simulation we then applied both current along $\hat{\mathbf x}$ and field along $\hat{\mathbf z}$. For the N\'eel DW in which $\mathbf m$ goes from $+\hat{\mathbf z}$ through $+\hat{\mathbf x}$ to $-\hat{\mathbf z}$ from left to right (see Fig.~\ref{fig:DW_sketch}), the current and field both drive the DW in the same direction (to the right). We observe that the solution quickly approaches a nearly 1D steadily propagating DW profile corresponding to the Walker type solution constructed above.

\textit{Conclusions.} We have studied the model of ultrathin
ferromagnetic film with interfacial DMI and two magnetic anisotropies.
When the out-of-plane anisotropy is stronger than the in-plane
anisotropy, we have found an exact 2D traveling wave DW solution
[Eqs.~(\ref{eq:walkersol}) and \eqref{eq:mwalker}] driven by both
electric current and magnetic field. This solution is an analog of
the well-known Walker solution for a 1D steadily moving DW. The
presence of an in-plane anisotropy is crucial to stabilize this
solution, and moreover it allows us to find analytical expressions for
the DW propagation direction and velocity [see Eqs.~(\ref{eq:nsign})
and~(\ref{eq:nn})] as functions of all material parameters.

\textit{Acknowledgments.} O.\,A.\,T. acknowledges support by the Grants-in-Aid for Scientific Research (Grants No.\,17K05511 and No.\,17H05173) from the Ministry of Education, Culture, Sports, Science and Technology (MEXT), Japan, MaHoJeRo grant (DAAD Spintronics network, Project No.\,57334897), by the grant of the Center for Science and Innovation in Spintronics (Core Research Cluster), Tohoku University, and by JSPS and RFBR under the Japan-Russian Research Cooperative Program. C.\,B.\,M. was supported by NSF via Grant No.\,DMS-1614948. V.\,V.\,S. and J.\,M.\,R. would like to acknowledge support from Leverhulme Trust Grant No.\,RPG-2014-226.

\bibliography{DW_references}

\end{document}